# Determination of localized conduction band-tail states distribution in single phase undoped microcrystalline silicon

Sanjay K. Ram[*,a], Satyendra Kumar[*,b] and P. Roca i Cabarrocas[!]
[*]*Department of Physics & Samtel Centre for Display Technologies, Indian Institute of Technology Kanpur, Kanpur-208016, India*
[!]*LPICM, UMR 7647 - CNRS - Ecole Polytechnique, 91128 Palaiseau Cedex, France*

We report on the phototransport properties of microstructurally well characterized plasma deposited highly crystallized microcrystalline silicon ($\mu$c-Si:H) films. The steady state photoconductivity, $\sigma_{ph}$, was measured on a wide microstructural variety of single-phase undoped $\mu$c-Si:H films as a function of temperature and light intensity. The band-tail parameter ($kT_c$) was calculated from the photoconductivity light intensity exponent ($\gamma$) values at different temperatures for a range of quasi-Fermi energies. The localized tail states distribution in the vicinity of conduction band edge of $\mu$c-Si:H was estimated using the values of $kT_c$. Our study shows that $\mu$c-Si:H films possessing dissimilar microstructural attributes exhibit different phototransport behaviors, which are linked to different features of the density of states maps of the material.

PACS numbers: 73.50.Pz, 73.50.–h, 73.61.–r, 71.23.–k, 73.50.Gr, 72.20.J

## 1. INTRODUCTION

Plasma deposited hydrogenated microcrystalline silicon ($\mu$c-Si:H) has immense potential in large area electronic applications [1,2] as it offers the possibilities of high carrier mobilities [3] and better stability against light and current induced degradation as compared with amorphous silicon (*a*-Si:H) [4,5]. However, the recombination mechanisms and the nature / distribution of the density of gap states (DOS) in $\mu$c-Si:H system are not completely elucidated due to the heterogeneous nature of the material, which is greatly influenced by the processing history [6,7,8,9]. For the same reason, a single DOS profile that is applicable for a certain microstructural type of material may fail to satisfy a different type of $\mu$c-Si:H.

Photoconductivity and its recombination kinetics are extensively used to study the effects of the density and the nature of the gap states in disordered materials [10]. In general, photoconductivity in disordered materials exhibits a non-integer power law dependence on carrier generation rate $G_L$ over several orders of magnitude given by:

$$\sigma_{ph} \propto G_L^{\gamma}, \tag{1}$$

[a] Corresponding author. E-mail address: skram@iitk.ac.in, sanjayk.ram@gmail.com
[b] satyen@iitk.ac.in

The photoconductivity light intensity exponent, $\gamma$ provides information about the recombination mechanisms in a semiconductor material. For $\gamma$ value lying between 0.5 and 1, Rose [10] described a model in which the density of trapped electrons $n_t$, approximated by

$$n_t(T) = \int_{-\infty}^{E_{fn}(T)} g(E)dE, \tag{2}$$

tracks the density of positively charged recombination centers $P_r$ to maintain charge neutrality. Rose showed that if the discrete states are distributed exponentially in the vicinity of band edges in the form of $e^{-\Delta E/kT_C}$, the photocurrent and light intensity curve should have the power $\gamma$ as:

$$\gamma = \frac{kT_c}{(kT + kT_c)}, \tag{3}$$

here $\Delta E$ is measured from the bottom of the conduction band (CB) and $kT_c$ is the characteristic energy of conduction band tail (CBT) *greater than or equal* to the energy corresponding to the measurement temperature *T*. In that case, most of the trapped electrons reside within $kT$ of the steady state Fermi level $E_{fn}$. The model as given by Eq. (2) is also applicable to the cases where DOS profile $g(E)$ does not decay in a purely exponential manner. The values for $\gamma$ then correspond to certain positions of $E_{fn}$, and an approximating exponential function with a local band tail parameter $kT_0(E_{fn})$ in its vicinity, giving $g(E_{fn})$.

The quasi Fermi level can be determined from the photoconductivity in the same range as was used to evaluate $\gamma$. The expression is given by:

$$\left[E_{\mathrm{c}}-E_{\mathrm{fn}}(\phi,T)\right]=\left[E_{\mathrm{c}}-E_{\mathrm{f}}(T)\right]-kT\ln\left[\frac{\sigma_{\mathrm{ph}}(\phi,T)}{\sigma_{\mathrm{d}}(T)}\right] \tag{4}$$

The above expression can be re-written in the way as below:

$$E_{\mathrm{c}}-E_{\mathrm{fn}}=kT\ln\left[\frac{\sigma_0}{\sigma_{\mathrm{ph}}(T)}\right] \tag{5}$$

Steady state photoconductivity (SSPC) is a technique widely used to study the gap states over a wide range in the bandgap. The SSPC technique is sensitive to both density and nature of all defect states acting as recombination centers between the quasi Fermi levels in the bandgap. In this article, we report the findings of our study of phototransport properties of microstructurally different $\mu$c-Si:H films using SSPC. Our results provide information about the localized states in the vicinity of CB edge.

## 2. EXPERIMENTAL DETAILS

We prepared a series of highly crystallized undoped μc-Si:H films having varying degree of crystallinity by depositing on Corning 1737 substrates at a substrate temperature of 200°C in a parallel-plate glow discharge plasma deposition system operating at a standard rf frequency of 13.56 MHz using high purity $SiF_4$, Ar and $H_2$ as feed gases. For the structural investigations, we employed variety of structural probes like in-situ spectroscopic ellipsometry (SE), Raman scattering (RS, from film and substrate side), X-ray diffraction (XRD) and atomic force microscopy (AFM) [11] These well-characterized films were studied for the electron transport behavior using dark conductivity and photoconductivity as functions of several discerning parameters such as temperature, wavelength and intensity of probing light. The effect of light intensity variation on the SSPC was probed using above-bandgap light (He-Ne laser, λ = 632.8 nm) in the temperature range of 20K–324K. Photon flux $\phi$ was varied from $\approx 10^{11}$ to $10^{17}$ photons/cm$^2$-sec using neutral density filters giving rise to generation rates of $\approx G_{\mathrm{L}}=10^{15}\text{-}10^{21}$ cm$^{-3}$s$^{-1}$. Penetration depth of this light is ~500 nm.

## 3. RESULTS AND DISCUSSION

The microstructural investigations using RS and SE demonstrated high total crystallinity of all the samples. SE results show a crystalline volume fraction >90% from the initial stages of growth, with the rest being density deficit in the bulk of the films. The bulk layer contains no amorphous phase. The detailed composition of the films educed from SE data shows grains of two distinct sizes, which is corroborated by the deconvolution of RS profiles using a crystallite size distribution of large grains (LG ~ 70-80 nm) and small grains (SG ≤ 6-8 nm). With film growth there is significant increase in the percentage volume fraction of LG. The XRD results showed presence of two sizes of crystallites and presence of preferred orientation in fully grown films. The commencement and evolution of conglomeration of crystallites with film growth demonstrated by AFM corresponds to the appearance and increase in the LG fraction [11].

The microstructurally different films were classified into three different types (*A*, *B* and *C*) based on the features of microstructure and growth type present and the corresponding co-planar electrical transport behaviors, to facilitate and systematize our study [12]. In the present contribution, we report the results of two samples: #B22 and #B23 representative of *type-A* and *type-B* $\mu$c-Si:H materials respectively. Outlining the microstructure of these two types briefly, the *type-A* films have small grains, low amount of conglomeration (without column formation), and high density of inter-grain boundary regions containing disordered phase. The *type-B* films show a rising fraction of LG. There is a marked morphological variation in these films due to the commencement of conglomeration of grains resulting in column formation, and a moderate amount of disordered phase is present in the columnar boundaries.

The results of temperature dependent photoconductivity for various light intensities, $\sigma_{\mathrm{ph}}(T,\phi)$ for *type-A* sample are shown in Fig. 1 (a). The temperature dependent dark conductivity $[\sigma_{\mathrm{d}}(T)]$ of this sample is shown by a solid line in Fig. 1 (a). The $\sigma_{\mathrm{d}}(T)$ shows an activated behavior over a large temperature range (≈170–450 K) with an activation energy $E_a$ = 0.5 eV. The $\sigma_{\mathrm{ph}}(T)$ of this sample is essentially an increasing function of $T$ for any value of light intensity used in the range mentioned above. However, at higher temperatures ($T$ = 324–175K), $\sigma_{\mathrm{ph}}$ is seen to decrease with increasing temperature, an effect known as thermal quenching (TQ). The power law behavior of the photocurrent of this sample with a change in photon flux was observed throughout the temperature range of our study. The temperature dependence of light intensity exponent, $\gamma(T)$ calculated at each measurement temperature from $\sigma_{\mathrm{ph}}(\phi)$ of the same sample is also plotted in Fig. 1(a) in the right side of Y-axis. The variation of $\gamma$ is found to be between 0.5 and 1 in the whole temperature range.

The $\sigma_{\mathrm{ph}}(T,\phi)$ for *type-B* sample is shown in Fig. 1(b), demonstrating a different behavior compared to the above case. $\sigma_{\mathrm{ph}}$ is found to increase monotonically with $T$ without any TQ effect. The solid line in this figure represents the $\sigma_{\mathrm{d}}$ with $E_a$ = 0.34 eV. The $\gamma(T)$ values obtained for different temperatures for this sample are plotted against reciprocal of $T$ and depicted in right side of Y-axis in Fig. 1(b). The $\gamma$ value of this sample is found to vary from 0.5 to 1 and it never goes down below the value 0.5 in the whole temperature range.

According to Rose model [10], when the $\gamma$ value lies between 0.5 and 1, it is assumed that the discrete states

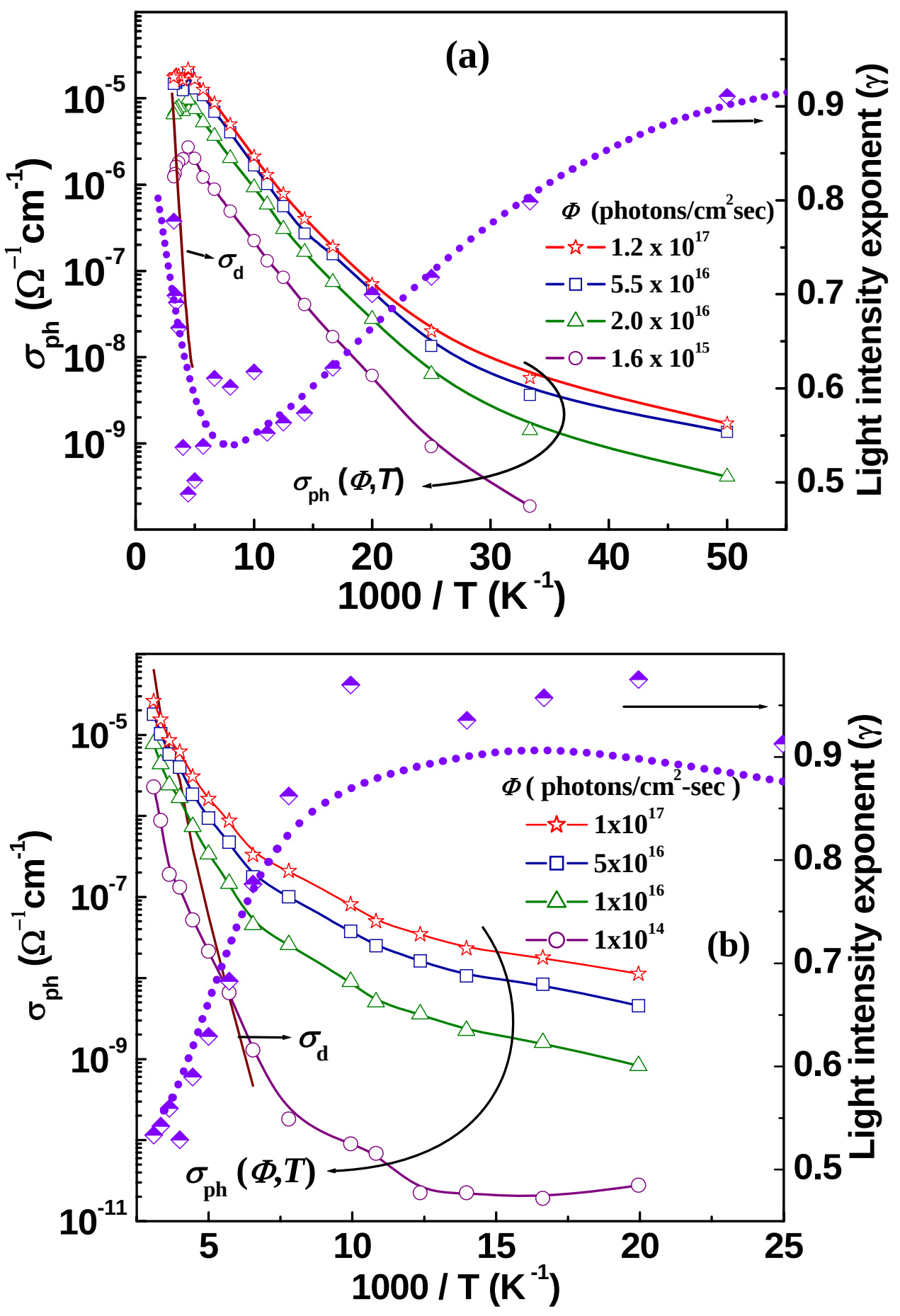

Fig. 1 The $\sigma_{ph}(\phi,T)$ (line+symbol) and $\gamma(T)$ (scatter points with dotted line to guide the eye) of samples of *type-A* (#B22) and *type-B* (#B23) are shown in parts (a) and (b) respectively. The temperature dependence of $\sigma_d$ of the samples are shown in respective graphs by solid lines.

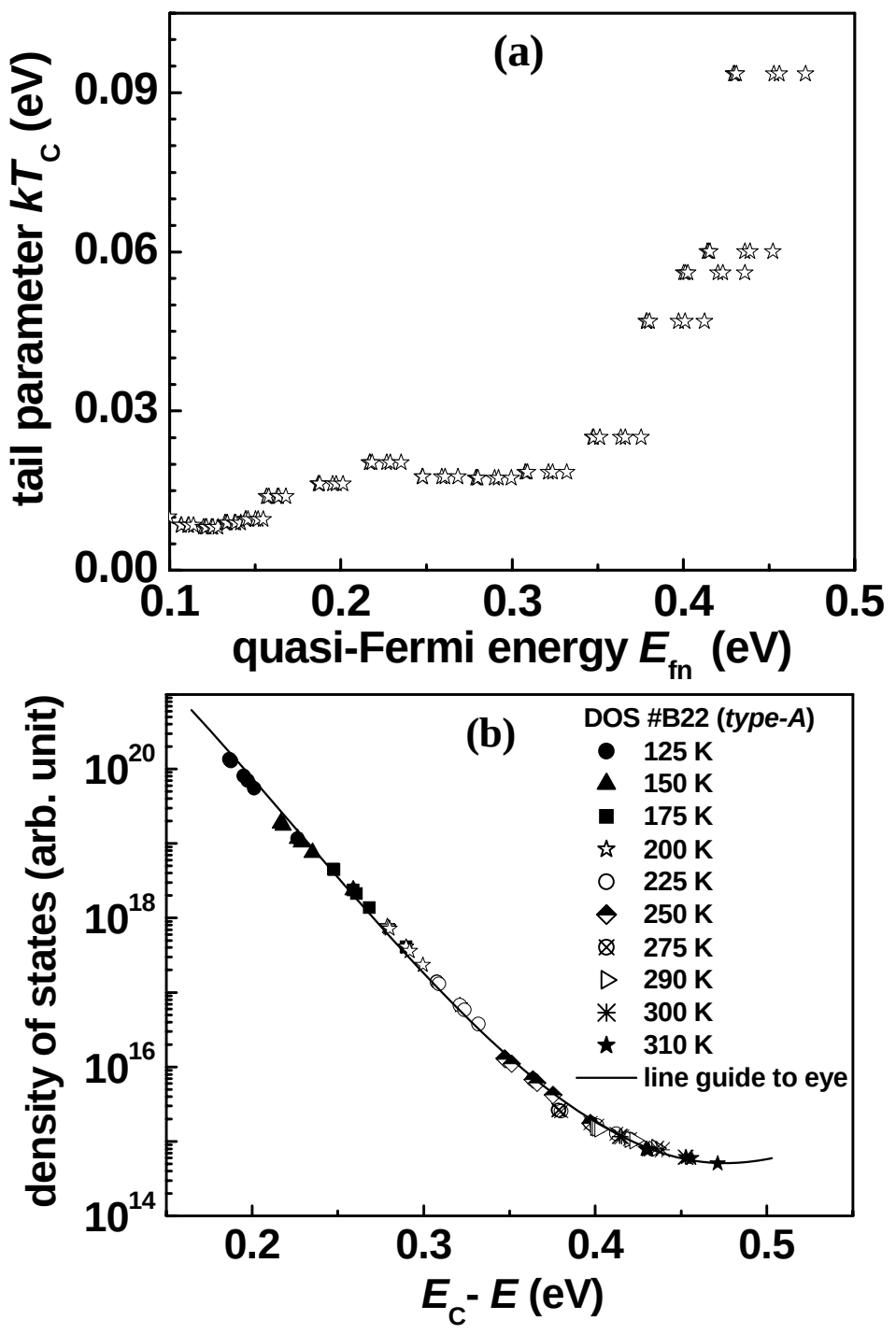

Fig. 2 (a) Plot of $kT_c$ vs. $E_{fn}$ of sample of *type-A* (#B22), where $kT_c$ was calculated from the $\gamma$ values at different temperatures shown in Fig. 1(a); (b) DOS distribution obtained by fitting $kT_c$ vs ($E_c$ -$E$) data of part (a) to exponential distribution of states. This sketch of DOS profile in this sample is only an approximation.

are distributed exponentially in the gap. We applied the model to the results of the phototransport studies of the *type-A* and *type-B* material shown in Fig. 1(a) and (b) respectively. The band-tail parameter $kT_c$ was calculated from the $\gamma$ values at different temperatures by applying the Eq. (3). The Fig. 2(a) depicts the calculated $kT_c$ of *type-A* material. The quasi Fermi level was determined using Eq. (4) and (5), from those photoconductivity values that lie in the range of a particular constant value of $\gamma$ for any particular $T$. The condition, $kT_c > kT$ is maintained in the whole temperature range of our study. The temperature range for which the data was evaluated in terms of the model in Eq. (5) varies from 324 K down to 128 K. Lower values of $T$ were not used in the calculation as hopping conduction might be operating in that range. Finally, based on the values of band-tail parameter (at different temperatures and for different energetic positions) obtained from Fig. 2(a), we calculate the density of states below the CB edge for this *type-A* material. The approximated DOS profile is shown in Fig. 2(b). Here, the numerical values have been used only to provide the characteristic energetic slopes with a visually appreciable structure of the DOS profile. It should be noted that although this DOS profile is only a rough estimate, but it can help us visualize the shape of the localized band tail distribution near one of the band edges, in this case, the CB edge. The density of localized tail states is found to be exponentially distributed in the CB region, though at deeper energetic positions the DOS decays very slowly.

Now let us turn to the results of the phototransport studies of the *type-B* material shown in Fig. 1(b). The method of calculation of band-tail parameter $kT_c$ and the quasi Fermi level at different temperatures was similar to the above case and the plot of values obtained is shown in Fig. 3(a). The condition $kT_c > kT$ is maintained in this case too. Here also low temperature region is not considered due to the possibility of hopping conduction operating in that range. The Fig. 3(b) shows a rough sketch of the density of states profile below the CB edge for the *type-B* $\mu$c-Si:H material. In this case, DOS is found to have slowly decaying states in the energy interval 0.2 to 0.3 eV, followed by a steeper tail. Similar to the DOS of *type-A*, the density of localized tail states at deeper energetic position shows less steep tail.

We see that in both the samples, the behavior of $\gamma$ is in agreement with Rose model, suggesting the presence of a band-tail state distribution in these two types of materials. For comparison we have plotted our experimentally estimated DOS profiles of both the types of $\mu$c-Si:H

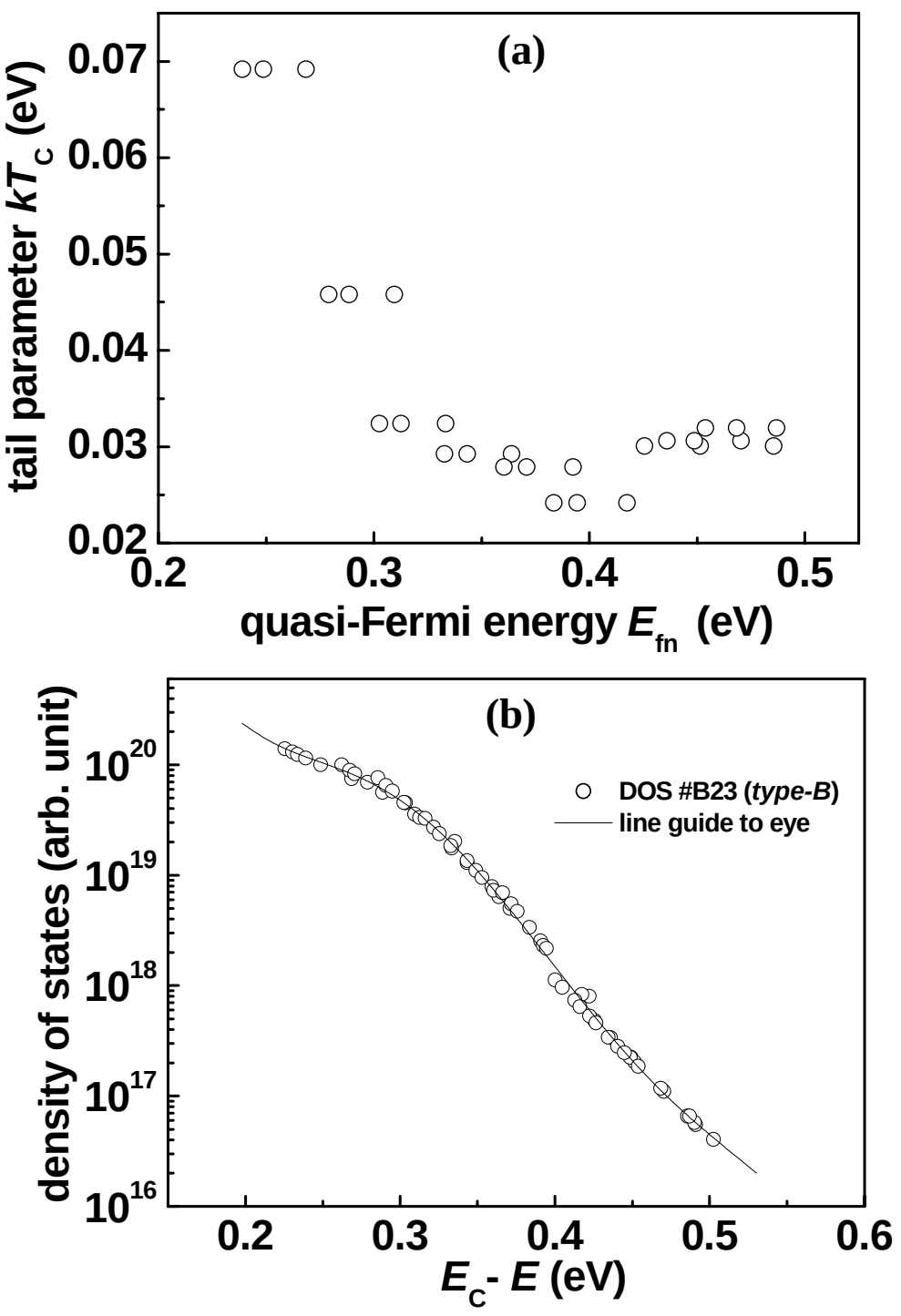

Fig. 3 (a) Plot of $kT_c$ vs. $E_{fn}$ of sample of *type-B* (#B23), where $kT_c$ was calculated from the $\gamma$ values at different temperatures shown in Fig. 1(b); (b) DOS distribution obtained by fitting $kT_c$ vs ($E_c$ -$E$) data of part (a) to exponential distribution of states. This sketch of DOS profile in this sample is only an approximation.

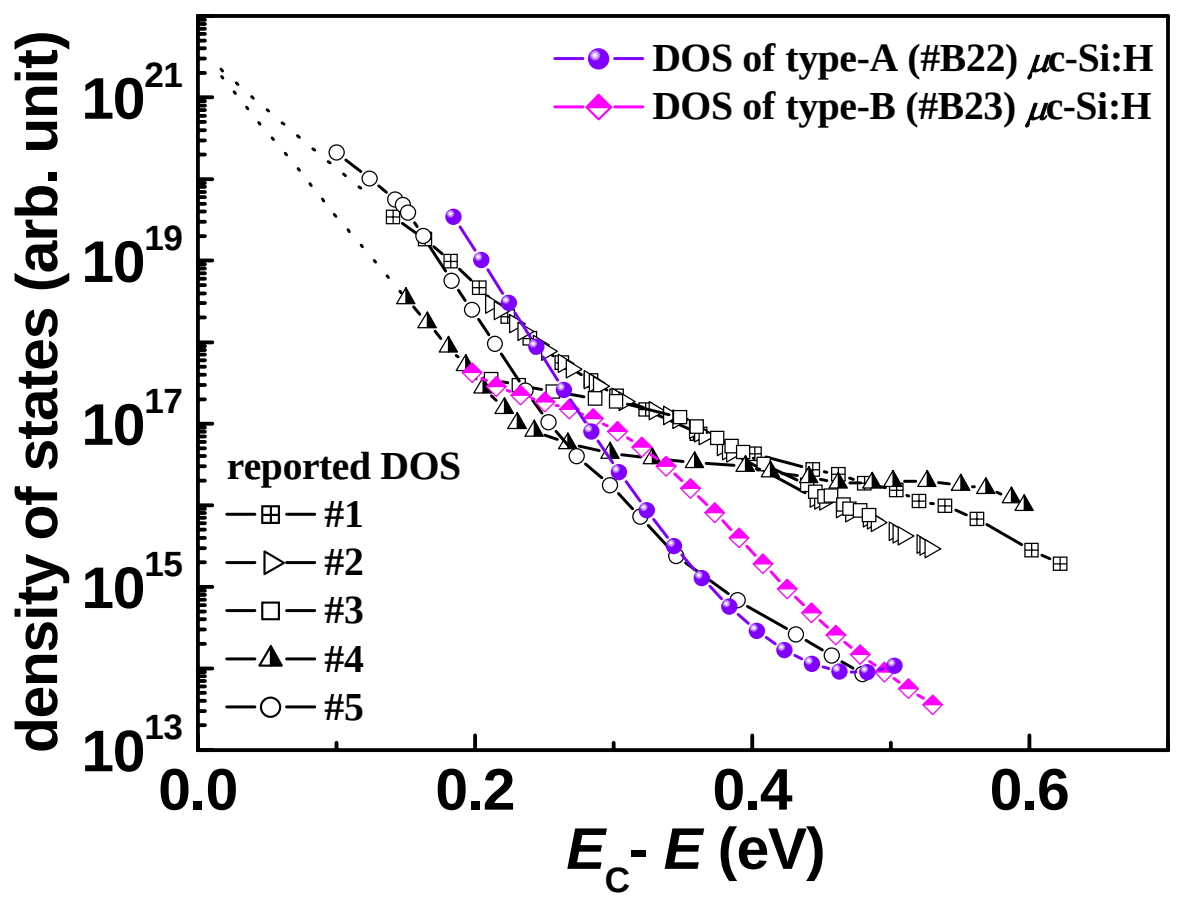

Fig. 4 DOS distribution obtained for SSPC measurement of *types-A* and *B* $\mu$c-Si:H samples are plotted along with DOS profiles of $\mu$c-Si:H suggested in literature obtained with other experimental techniques (#1-5). Here, #1 [Ref. 8; MPC-DOS coplanar $\mu$c-Si:H ($I_{crs}$=0.5)], #2 [Ref. 5,9; MPC-DOS HW-CVD $\mu$c-Si:H], #3 [Ref. 5,9; MPC-DOS SPC $\mu$c-Si:H], #4 [Ref. 7; TOF-DOS $\mu$c-Si:H], #5 [Ref. 6; SSPC-DOS $\mu$c-Si:H].

along with the DOS profiles of $\mu$c-Si:H suggested in the literature derived from other experimental techniques (#1-5) [5,6,7,8,9] in Fig. 4. The obtained DOS profiles of our $\mu$c-Si:H samples are in good agreement with the DOS described in these studies [5,6,7,8,9]. The shape of our DOS profile of *type-A* material is quite similar to the SSPC-DOS estimated by Bruggemann (#5) [6]. However, the shape of DOS profile of *type-B* material near the band edge (from 0.2 to 0.3 eV) is similar to that of DOS obtained by modulated photoconductivity (MPC-DOS) in solid phase crystallized (SPC) $\mu$c-Si:H (#3) [5,9]. The reason could be due to the highly crystalline nature of our *type-B* material. The lower values of DOS of our $\mu$c-Si:H at deeper energetic position in the gap can be the result of the passivation of grain boundaries by $H_2$, while the boundaries of SPC $\mu$c-Si:H material are not passivated, and can therefore give rise to high value of DOS at deeper energies in the gap. In order to understand the reason behind the phototransport behavior in our material, we must proceed by considering the microstructure of the films. It can be recalled that *type-A* $\mu$c-Si:H films are crystallized with small grains. The smaller grain size results in a higher number density of boundary regions, the unsaturated DBs are assumed to be located in these disordered boundary regions. Therefore, the material may possess quite a significant number of DB densities, though less than what is present in *a*-Si:H, but enough to result in a TQ effect. In addition to the presence of DBs, the asymmetry in the band tail states is also a known reason behind *TQ* effect in *a*-Si:H [13]. In our $\mu$c-Si:H films, the results of the dependence of sub gap absorption on varying microstructure suggest that the width of VBT of *type-A* material should be larger than *type-B* material. The VBT width of *type-A* material is closer to that of *a*-Si:H and larger than the width of CBT, giving rise to asymmetry in band tail states in *type-A* material. Another important aspect is the absence of superlinear behavior of $\gamma$ in *type-A* material, which typically accompanies TQ in *a*-Si:H. The dark conductivity results of this sample predict the position of $E_f \approx (E_c$-0.46) eV, which is at a higher position than that found in undoped *a*-Si:H. This may be the reason behind the absence of superlinear behavior.

Now we come to the phototransport properties observed in *type-B* material. The absence of *TQ* emphasizes that both the tail states should be somewhat symmetric to each other. A rough sketch of the effective DOS profile in the vicinity of CB edge suggests that instead of a single exponential band tail of CBT, some hybrid form of DOS is approximated. From this information we can infer a similar possible shape of DOS near the VBT in *type-B* material. This has been demonstrated by our studies of numerical modeling of SSPC [14].

## 4. SUMMARY

The results of phototransport studies of our highly crystallized undoped single-phase $\mu$c-Si:H films indicate that in the heterogeneous $\mu$c-Si:H system, films of the same material having different microstructures can have different phototransport behavior, different underlying mechanisms and different effective DOS distributions.

**ACKNOWLEDGEMENTS**

One of the authors (SKR) gratefully acknowledges Dean of R & D, I.I.T. Kanpur, Samtel Centre for Display Technologies, I.I.T. Kanpur, Council of Scientific and Industrial Research, New Delhi and Department of Science and Technology, New Delhi, for providing financial support.